# Dynamic full field optical coherence tomography: subcellular metabolic contrast revealed in tissues by temporal analysis of interferometric signals

Clement Apelian*[a,b], Fabrice Harms[b], Olivier Thouvenin[a] and A. Claude Boccara[a,b]

[a]Institut Langevin – ESPCI ParisTech, , PSL Research University 1 rue Jussieu, PARIS, France, 75005; [b]LLTech SAS, Pépinière Paris Santé Cochin 29 rue du Faubourg Saint Jacques, PARIS, France, 75014

## Abstract

We developed a new endogenous approach to reveal subcellular metabolic contrast in fresh ex vivo tissues taking advantage of the time dependence of the full field optical coherence tomography interferometric signals. This method reveals signals linked with local activity of the endogenous scattering elements which can reveal cells where other imaging techniques fail or need exogenous contrast agents.  We benefit from the micrometric transverse resolution of full field OCT to image intracellular features. We used this time dependence to identify different dynamics at the millisecond scale on a wide range of organs in normal or pathological conditions.

## 1. Introduction

Life, from the simplest unicellular being to complex organisms, is an astonishing entanglement of multiscale structures and dynamics[1]. The dynamic properties of subcellular structures are important markers of health[2].

A number of methods have successfully been used to explore these dynamics, such as fluorescent correlation spectroscopy (FCS)[3,4]. This technique records intensity fluctuations of fluorescent particles. Autocorrelation assisted models are then used to extract quantitative information about diffusion dynamics within the sample volume. Phase contrast microscopy (PCM) records phase shifts (i.e. optical path length variations) to reveal contrast that is invisible in standard microscopy. The motion of various subcellular components have successfully been recorded with PCM to study cellular motility[5,6]. Another approach that has been developed for study of motion such as flow or diffusion of particles relies on optical coherence tomography (OCT). Named dynamic light scattering OCT (DLS-OCT), it uses autocorrelation to extract quantitative information such as axial velocity and diffusion based on mathematical models[7].

Imaging the dynamics of cells is an effective way to test their response to drugs or pathogens; however cellular properties often change when shifting from 2D cultures to tissues. More sophisticated models have therefore been developed such as scaffolds imaged with OCT[8], and spheroids imaged with holography[9], for instance. The increase in complexity on passing from 2D cell cultures to 3D tissue structures presents new challenges for imaging and recording dynamics.

Full field OCT method (FFOCT) enables imaging through the depth of tissues, reaching depths of hundreds of microns even in highly scattering tissues[10] with an isotropic resolution of 1µm or less, i.e. up to a factor of  ten times better than OCT. This endogenous, non-invasive and non-destructive technique has been used to detect cancers with specificity and sensitivity values comparable to traditional techniques such as frozen section [11,12]. However individual cells and nuclei are not always visible in FFOCT but are often crucial to the pathologist's diagnosis. Highly backscattering structures (e.g. connective tissue such as collagen) dominate the FFOCT signal, masking low or equivalent backscattering structures such as cells and their nuclei.

The use of contrast agents like dyes, fluorophores or nanoparticles[13] is often helpful to reveal cells. However, introducing exogenous agents inside the medium could be problematic if further chemical or biological (DNA) pathological testing is needed or in if the tissue is to be conserved.

Using the time dimension to provide supplementary information about the tissue dynamics has already been shown[14–16] using OCT. However due to the limited resolution of OCT (around 5 to 15μm) cellular or sub cellular information was scarce. We propose here to take advantage of the higher resolution of FFOCT in order to probe dynamics at sub cellular level. We perform parallel acquisition of time signals within the coherence gate (corresponding to the selected layer in the sample volume)[10]. Our technique is based on two technical improvements that change the scale of dynamics we are sensible to. We use higher resolution than OCT to access intracellular contrast which allows considering cells' state. We also record at frame rate higher than other dynamic studies to capture rapid motion of biological scattering elements that would be missed with slower image acquisition rate. While OCT can observe long term phenomena like cell apoptosis, proliferation or migration[8,16], we looked here at subcellular behaviour at short timescales (from <10 ms) with negligible cellular migration or reorganization. Long-term acquisitions similar to those performed in OCT could also be achieved by extending the acquisition time or by time lapse imaging[17].

In this paper we describe a new approach based on the time dependence of the FFOCT signal. The additional contrast in dynamic FFOCT (D-FFOCT) is provided by metabolic activity, in situ, at subcellular scale.

## 2. Principle of D-FFOCT

D-FFOCT is based on the usual FFOCT setup[9] (figure 1). It consists of a Linnik interferometer with incoherent illumination where one arm contains the sample to image. The backscattered light from each voxel at the chosen depth (i.e. a layer orthogonal layer to the optical axis) interferes with the light of each pixel of the reference arm and is imaged on the CCD camera. In standard FFOCT, the interferometric amplitude is obtained by combination of images for which a modulation of the path difference is induced in the reference arm. In D-FFOCT, there is no reference modulation and only movements in the sample itself contribute to the signal modulation. More precisely, we record a time dependent interferometric backscattered signal superimposed on a stationary signal:

$$I(x, y, z, t) = I_0(x, y, z) + \alpha(x, y, z, t) \cdot \cos(\phi(x, y, z, t)) \tag{1}$$

where I is the intensity received on the camera, $I_0$ comes from the reference signal and the stationary backscattered light inside and outside of the coherence gate, α is the amplitude and ϕ the phase of the backscattered signal interfering with the reference..

For a given duration t we work at a fixed depth z and set the coherence layer (coherence gate) to be at the centre of the depth of field of the objective. We can move the sample or the interferometer to scan the z axis and/or to explore wider fields in the (x,y) plane. One can thus reconstruct 3D volumes and/or wide fields of view by stitching different layers or fields together.

Temporal analysis is used to extract useful features in the time-dependent FFOCT images. $I_0$ is not time dependent and therefore can be subtracted as a continuous term. One is then left with a signal corresponding to motion of scatterers. This signal is a function of both amplitude and phase. The amplitude, α, is linked to the size, the refractive index and number of scattering structures inside the voxel for a given time. The phase, ϕ, is determined by the spatial distribution of scatterers, and this distribution is changing due to the motion of the scatterers inside the voxels. Data processing extracts meaningful information by calculating the standard deviation of the signal in time. This suppresses the signal from highly scattering stationary structures such as collagen or myelin fibres. Strongly backscattering structures can dominate the signal even outside of the full width half maximum (FWHM) of the coherence gate, thereby masking weakly scattering structures such as cells. The amplitude of the backscattered signal from weakly scattering structures is down to 100 times lower than the amplitude of strongly scattering structures, or $10^4$ times lower in scattered intensity. However, scattering structures that happen to display the most dynamic behaviour are revealed by D-FFOCT.

To illustrate how the tissue dynamics impact the signal, one can consider the case of a single scattering particle moving uniformly along the vertical (i.e. optical) axis. Equation (1) shows that $I_0$ and $\alpha$ are constant in time if the particle remains inside the voxel. In this case a linear z motion in the voxel results in a sinusoidal signal I of amplitude $\alpha$ and offset $I_0$ (figure 1). For a group of scatterers moving together as a block (i.e. with no relative displacements between scatterers) the behaviour is similar along the vertical axis. If there is relative (tree dimensional) motion between scatterers, which is indeed the case due to Brownian motion of subcellular structures, we can decompose the phase as a sum of a global and a relative component:

$\phi(t) = 2\pi(\Delta(t)+\delta(t))/\lambda$   (2).

With $\Delta(t)$ the global displacement of the scatterers along the optical axis and $\delta(t)$ the effective distance that corresponds to phase variations introduced by the relative displacements between scatterers.

If the variations of $\delta(t)$ are fast in relation to the capture time T of the camera and if $\Delta(t)$ varies slowly compared to T then the recorded value of $\delta_T(t)$ on the camera will be negligible compared to $\Delta_T(t)$ and so the global motion will dominate.

As mentioned before, the other source of variability in the signal arises from $\alpha$. When a scatterer enters or leaves a voxel, the overall scattering properties vary. The envelope of the fluctuation changes will depend on the number of scatterers and on their scattering properties.

By analysing both contributions to the dynamic signal we can show that axial (z) sensitivity to displacement is larger than the transverse sensitivity (x,y). Indeed, the minimal z motion corresponding to a change from the minimum to the maximum of the $\cos(\phi)$ is $\lambda/4n$ (111 nm in our case, with a refractive index of n=1.4 and a central wavelength of $\lambda_c$=625 nm). In the x and y directions however, a scatterer has to move from one voxel to its neighbour to affect the signal, typically corresponding to half the point spread function (PSF) width (750 nm).

It has been shown, for instance, that mitochondria in liver display constrained Brownian motion and/or directed Brownian motion with occasional jumps[18]. These behaviours could be related to the time signals recorded on the camera for liver tissue. The signal presented in Figure 1_A might be the result of a scatterer jumping inside the voxel thus changing the envelope of the time fluctuation dramatically between two frames. In figure_1_B a constant increase is visible and could correspond to an increase of scatterers in the voxel during the acquisition time whereas in figure 1_C the fluctuations keep a constant mean value. Filtering out the highest frequencies from the figure 1_C signal, a quasi-periodic signal is revealed which corresponds roughly to the case considered previously of a global motion along the optical axis.

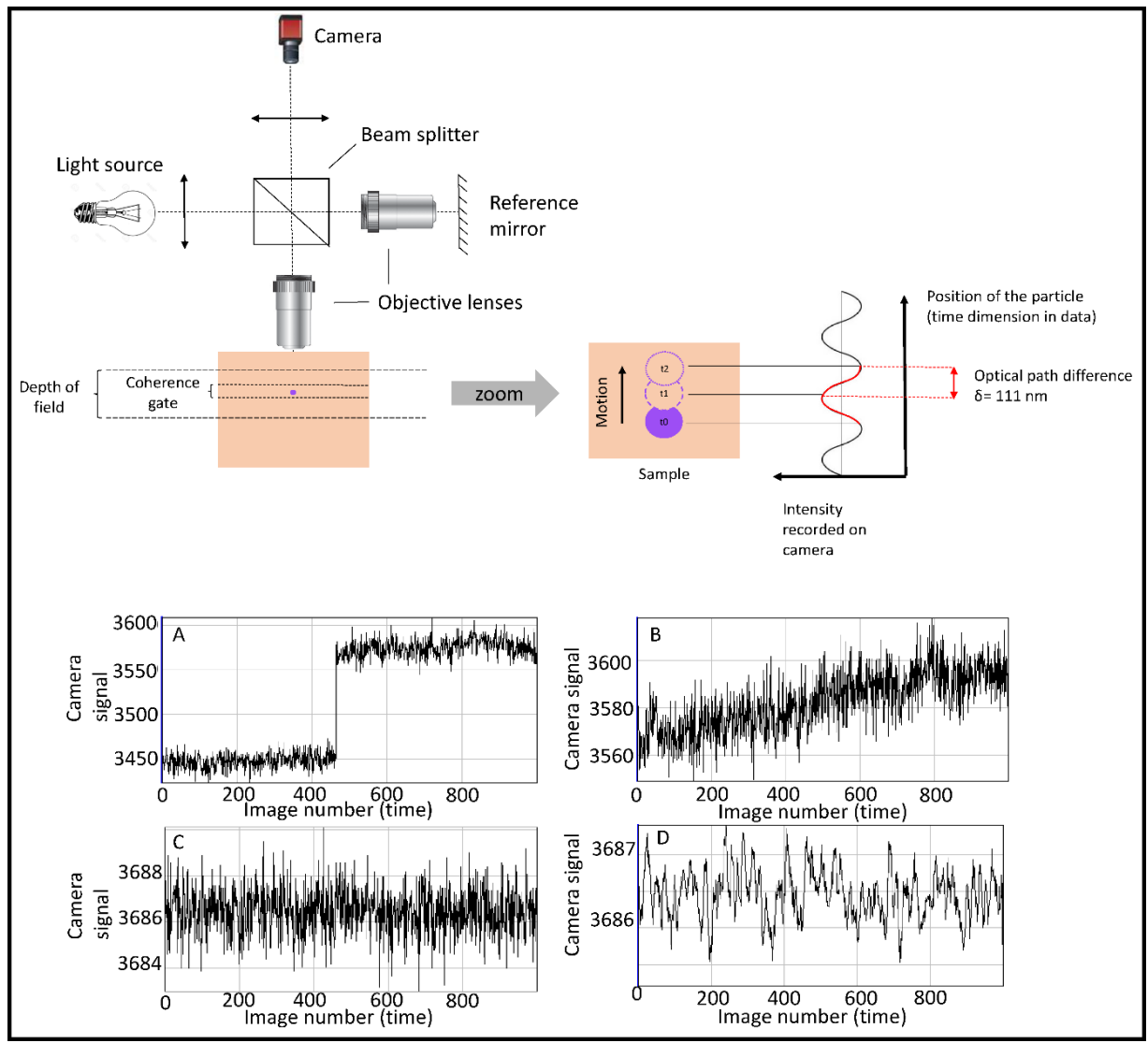

Figure 1: An FFOCT setup and illustration of the effect of a vertical movement of a single scatterer inside a voxel in terms of intensity variation on the camera. (A), (B) and (C) are time signals of three pixels of the same image of liver, (A) presents an important and sudden increase of scatterers while (B) seems to receive a continual flux of entering scatterers or a reorganization of scatterers present. (C) displays an equilibrium (balance between scatterers entering and leaving the voxel), (D) is (C) filtered by a low-pass filter to remove noise, where we can distinguish a quasi-periodic dynamic for that pixel with a period of approximately 150 ms. (One time step on the horizontal axis corresponds to 7 ms)

## 3. Material and methods

### 3.1 Experimental setup

Two setups were used. The first was a custom setup based on the commercially available Light-CT scanner (LLTech SAS, France). A second D-FFOCT setup was built in the laboratory to perform higher resolution imaging of activity within cells.

Pairs of 10x or 30x water immersion objectives with numerical apertures of 0.3 or 0.8 respectively were used. The frame rate of the camera limits the speed of dynamics one will be able to capture: considering that at least two frames are necessary to observe a movement, the setup currently used can reach 138Hz. The full well capacity determines, along with the power of the light source, the exposure time required to use the full dynamic range of the camera (200 000 $e^-$)[9]. The spectral width of the source determines the axial resolution and hence the thickness of the tissue slice contributing to the signal. The Light-CT

Scanner has a 1 μm axial resolution. The numerical aperture of the objectives and the central wavelength of the light source determine the transverse resolution which is set by the point spread function (PSF). The maximum depth we can access is tissue dependent. With the light source power and camera full well capacity used, we typically reach depths of 50 μm to 100 μm. It was important to ensure the mechanical stability of the interferometer. Since D-FFOCT is motion-based, any global movement will decrease the contrast and/or introduce stray signals. Stability was optimized by having both interferometer arms and the sample mounted in one single mechanical structure, thus suppressing leverage arms and moving parts. By measuring fringe drift over lengthy image acquisition durations (20 seconds), we recorded only 16 nm drift.

### 3.2 Processing

Raw data from the camera were processed to reveal images. FFOCT uses 2-phase or 4-phase step processing to produce the amplitude image[10], while dynamic light scattering uses decorrelation time[7,16] and motility-sensitive OCT uses the motility ratio[14,15]. We chose to base our processing on the standard deviation (STD) [14,15] as follows :

$$D(x,y) = < \sqrt{\frac{1}{N}\sum_{i=1}^{N}(S(x,y,t_i) - <S(x,y)>)^2} > \qquad (3)$$

where $D(x,y)$ is the D-FFOCT signal at a given pixel (x,y). The mean value of the standard deviation was calculated on a number of sub-stacks N of the raw movie S. The stack size (in nuber of frames, wich is equivalent to discret time) can be varied: the smaller the stack, the more efficiently "slower dynamics" are cut off. S was therefore separated into different frequency bands and $D(x,y)$ calculated for each frequency band to reveal specific dynamics. In this way, one cut-off can be used to separate two cell types for instance. Decorrelation maps have been tested to render dynamic contrast. We found however that collapsing the autocorrelation function of a pixel into one representative value such as the half width at half maximum does reveal a contrast that is poorer than STD processing contrast.

### 3.3 Sample ethics

Animals were housed individually with free access to food and water and maintained in a temperature-controlled environment on a 12 hour light/dark cycle. Experiments were performed in agreement with the institutional guidelines for the use and care of animals and in compliance with national and international laws and policies (Council directives no. 87-848, 19 October 1987, Ministère de l'Agriculture et de la Forêt, Service Vétérinaire de la Santé et de la Protection Animale). We used rat and mouse organs rapidly after dissection in order to image them while activity is still high inside the tissues (before apoptosis). A wide range of tissues were studied, including brain, liver, kidney, spleen, skin, muscle, heart, pancreas and intestine, in mouse and rat.

For human samples, informed and written consent was obtained in all cases, following the standard procedure of the concerned hospitals.

## 4. Results and discussion

With our dynamic approach we were able to show a wide range of new features that were not visible with FFOCT: thanks to the metabolic activity within cells, hidden cells were revealed and the overall visibility of tissue morphology is improved compared to OCT. The origin of the motion was studied and revealed to be induced by motility. We also identified different dynamics corresponding to different structures and scales and finally focused on the application of this technology to cancer diagnostics. Results were obtained on fresh ex vivo tissue from various organs, imaged shortly after sacrifice and dissection.

### 4.1 Dynamic versus static contrast on fresh excised tissue

In FFOCT images of brain tissue, myelin surrounding axons appears as the most scattering structure. Myelin is abundant and prevents us from imaging the (weakly backscattering) neurons (which appear in black in Figure 2_A1). The FFOCT signal backscattered from inside a cell is just above the noise floor, i.e. much lower than the signal from myelin. D-FFOCT imaging (Figure 2_B1) of cortical tissue almost completely erases the myelinated structures that are stationary and reveals cell bodies. These high activity areas are colocalized with some of the dark circular spots on the FFOCT images. Those that remain dark in D-FFOCT may be dead cells or otherwise inactive structures (e.g. capillaries) and thus produce no signal.

Subcellular metabolic activity is visible with D-FFOCT in large cells such as hepatocytes. An optically sectioned liver tissue generates a roughly homogenous FFOCT response and cell borders are often impossible to identify. On figure 2_A2 the FFOCT image of intact rat liver (whole organ, uncut) reveals the fibres of the capsule while the difference of activity between the cytoplasm and the connective tissue clearly indicates the border of the cells on the D-FFOCT image C2 (figure 2). Moreover, if we zoom in on the cell cytoplasm we can observe heterogeneities that are linked to different dynamics. These heterogeneities are localized in space and time. That means that 3D imaging of these tissues is possible. Even if imaging is done layer by layer, the areas of high activity are located in a defined 3D environment during the time of acquisition, which means that cells are well reconstructed in depth without artefacts introduced by overall motion of the cells (Figure 2_A3 and 2_B3). Large structures such as tubules in kidney (figure 2_C3) are most readily recognized in 3D reconstruction where tubules are seen as pipe-like structures instead of slices. These two tissue examples show that D-FFOCT adds complementary contrast to FFOCT by revealing cellular and sub-cellular features.

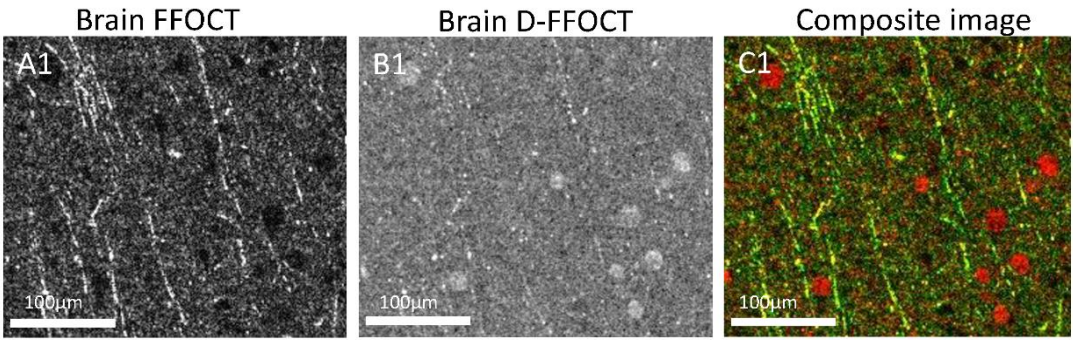
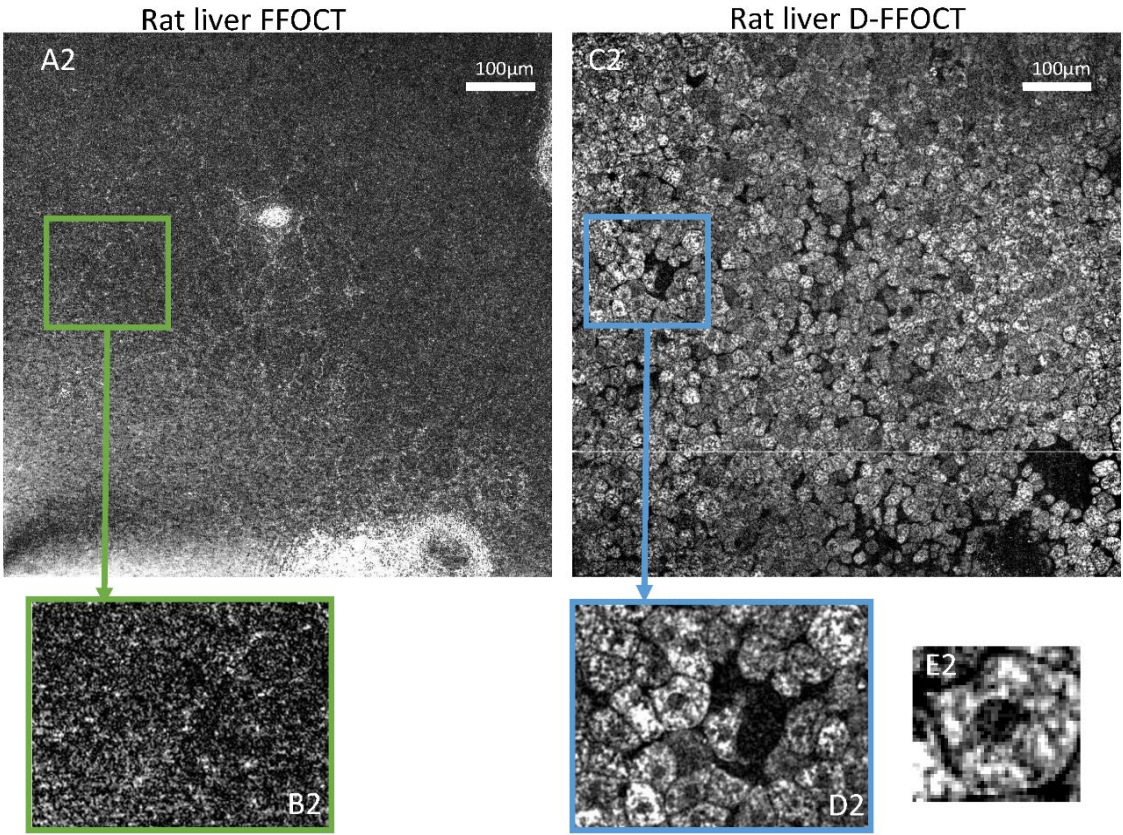
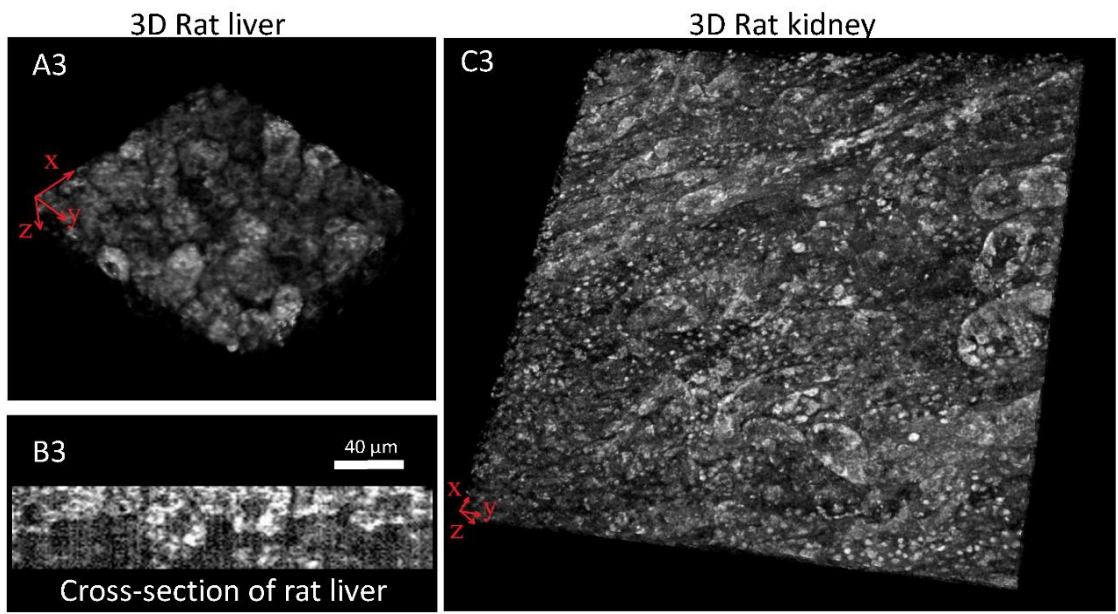

Figure 2: (A1) FFOCT of mouse brain cortex, (B1) D-FFOCT image of the same region and (C1) composite image shows superposition of the somas with OCT dark spots and axons bundles. Field of view 350x320µm.

FFOCT image of a rat liver (A2) with (B2) zoom on (A2), while (C2) and (D2) show the corresponding zones of the D-FFOCT image. (E2) is a zoom focused on one specific cell, we can observe different clusters of scatterers inside. This image is located on the surface of the rat liver with part of the capsule (highly reflective in A2) remaining in the lower part of the field. We can note that the D-FFOCT signal is weaker in those areas corresponding to where the capsule is thicker. (A2) and (C2) are 800x800µm, (B2) and (D2) are 325x250µm and (E2) is 65x65µm.

(A3) is a 3D D-FFOCT image of rat liver 170x170x51µm$^3$. Cells can be reconstructed along the z dimension and conserve a continuity in dynamic contrast. This means that the source of the contrast is moving in a confined volume and that no cell migration is occurring during the acquisition. (B3) corresponds to A3 in cross-section where cells are recognized without artefacts (image dimensions are 123x43µm$^2$). (C3) is a 3D view of kidney 604x604x42µm$^3$ where tubules are visible over the whole field, orientated from the lower left corner to the upper right one. We can also identify a lot of small objects that are yet to be identified.

### 4.2 Analysis of cellular activity

The D-FFOCT signal recorded is based on the motion of the scatterers. This motion could either be caused by motility or only Brownian diffusion with constraints. We first Fixed tissues and observed no D-FFOCT signals. Then a sample was imaged over several hours' duration and a signal loss over time was recorded (Figure 3). From this, we derived the hypothesis that as the signal from the cells was decreasing to reach the noise floor, the signal must be directly or indirectly caused by motility. . Our second hypothesis was that little oxygen was likely to be kept and consumed for that long inside an excised tissue, thus it was unlikely that normal energy production (i.e. cellular aerobic respiration) was the main source of energy in the cells. Finally, as liver cells have a D-FFOCT signal that lasts much longer than in other tissues, it is possible that high reserves of glucose could maintain a high and long-lasting dynamic signal if the main energy source of the cells was glycolysis. We tested these hypotheses with two experiments. First we made sure that the normal energy source of the cells from aerobic respiration was inhibited and then we tested glycolysis. To achieve this, we introduced an inhibitor of the metabolic chain we wanted to test, then we recorded the D-FFOCT signal during the action of the inhibitor. For the aerobic respiration we used rotenone as an inhibitor of complex I (NADH-ubiquinone oxydoreductase) The D-FFOCT signal was not significantly reduced.We therefore concluded that, as expected, aerobic respiration is not the main source of energy in our samples. We then used 2-deoxy-D-glucose to inhibit glycolysis and stop the energy production of the cells (Figure 3). The global contrast of the image dropped from 0.6 to 0.24 (corresponding to the noise floor, i.e. zero activity) in 30 minutes, which corresponds to the time the inhibitor needs to diffuse inside the sample volume. For the control group, this drop in activity to noise level takes 27 hours. The D-FFOCT signal is therefore related to glycolysis and thus local activity.

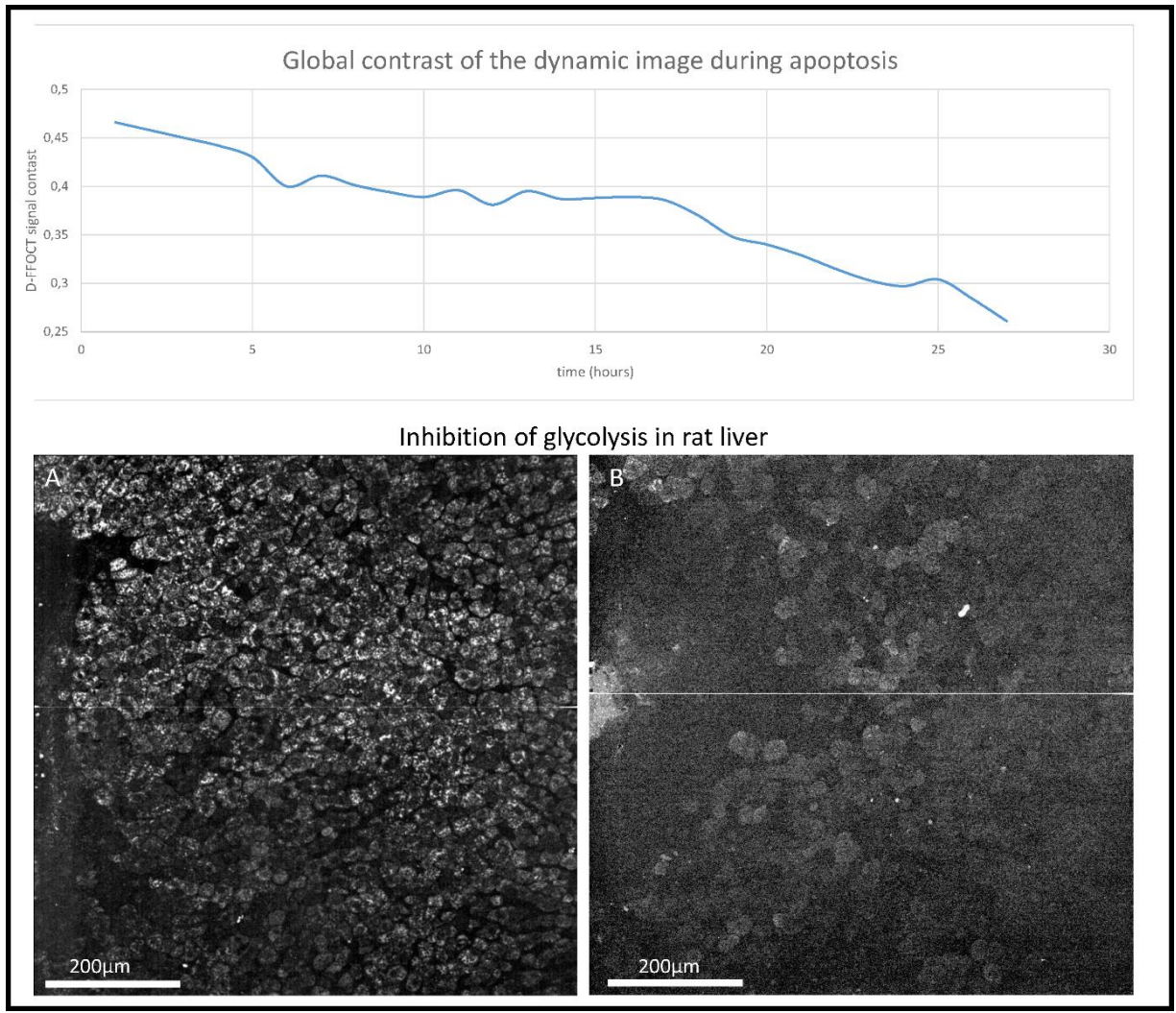

Figure 3: graph shows a decrease of the overall D-FFOCT contrast, the sample (rat liver) was imaged once an hour for 27 hours until it reached the noise level (typical signal level of the background tissue outside the cells = 0.25).

(A) D-FFOCT image of rat liver. An incision was made in the tissue to introduce the inhibitor (arrow). (B) is the same field of view after introducing the glycolysis inhibitor (2-deoxy-D-glucose). The inflammation has narrowed the incision and we see a clear drop in signal value (image B is multiplied by ten).

### 4.3 Multi-scale dynamics

We have shown that the D-FFOCT signal is linked to cell activity. If we could relate measurable differences in the time dependant signal to cell type or pathology, this could be of importance for diagnosis. In order to do this, we split the stack of data in time into smaller stacks of images and applied the dynamic processing. In this way, we can for instance identify specific dynamics inside different cells by looking at the autocorrelation function (Figure 4 graph, logarithmic scale). Signals whose autocorrelation functions decrease rapidly present rapid fluctuations, like erythrocytes, whereas those presenting a persistent autocorrelation signal, like hepatocytes in liver tissues, are related with slower dynamics. This observation is consistent with our knowledge of erythrocytes and hepatocytes. Erythrocytes are known to display rapid membrane fluctuation that could be the source of the dynamic contrast[6,19]. We improved this sorting by separating for different dynamic frequency bands. The typical threshold was set with an averaging window of 10 images (corresponding to 72 ms). We obtain two images at the end of this process, one rapid motion image corresponding to the high frequencies and a longer time-scale image corresponding to low frequencies in time signals. These images display complementary contrast, for example we can identify capillaries on the longer time-scale image and the erythrocytes within the capillaries on the rapid-motion image. We could also separate the fast scatterers inside an hepatocyte from the other slower-moving neighbours, thereby identifying different dynamics inside a single cell.

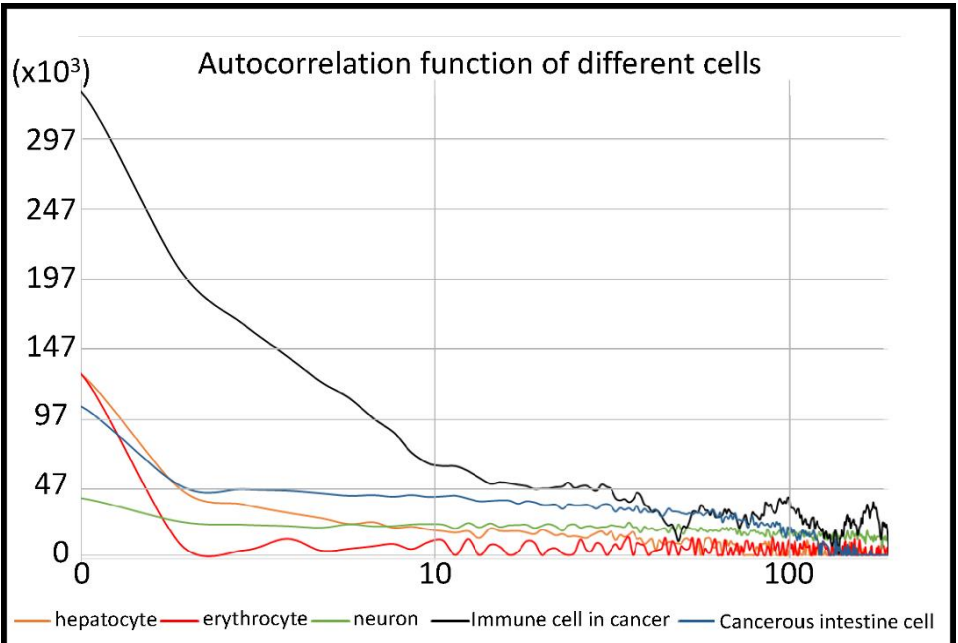

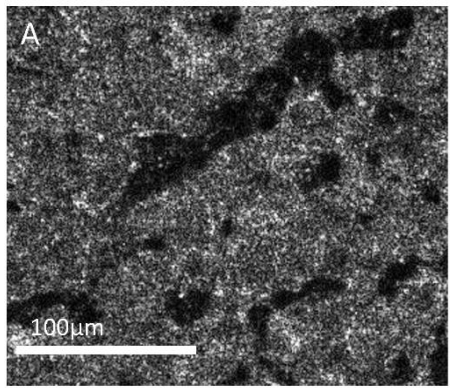
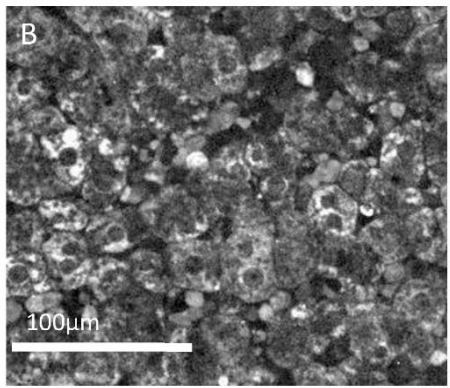

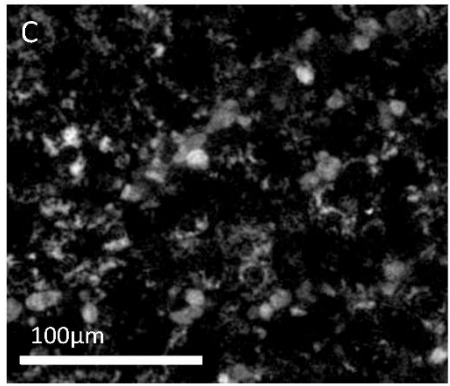
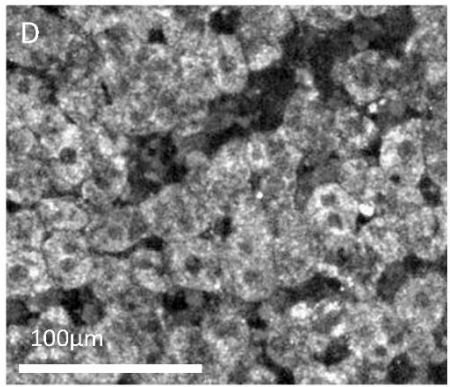

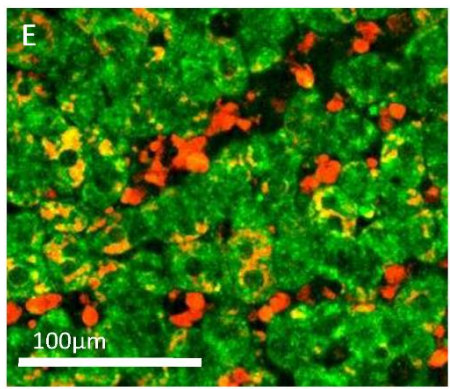
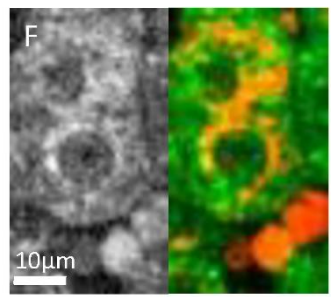

Figure 4: In this graph of the autocorrelation function of 5 different cell types in liver, brain and intestinal tissues. The dynamics of these cells are fundamentally different. We have the erythrocyte cell (in red) with the same signal energy (autocorrelation at the origin is the integral of the squared signal which is the energy) as the hepatocyte (orange) but with rapid fluctuations, while the hepatocyte decorrelates over a longer period. The neuron (green) takes even longer to decorrelate and has a lower energy. In cancerous regions of the mouse intestine, two cell types were identified: one with standard energy and long decorrelation times (blue) corresponding to cancer cells, and the other corresponding to immune cells (black), the highest D-FFOCT signal recorded, which also have long decorrelation times, but with a different damping factor. The noise signal has been removed from these curves.

Images (A-F) illustrate the discriminating power of multi-scale D-FFOCT analysis. (A) FFOCT image of rat liver, where the dark areas are capillaries and hepatocytes appear uniformly grey. (B) is the same field in D-FFOCT, where hepatocytes and erythrocytes are visible and there are inhomogeneities in the cytoplasm of hepatocytes. (C) and (D) are filtered data for respectively high and low frequencies. (E) is the overlay of (C) and (D) in red and green respectively. (F) is a zoom of the same field of (B) and (E).

### 4.4 Observations in cancerous tissues: towards an endogenous dynamic marker?

One important application of this imaging technique could be to help pathologists to diagnose cancer through careful analysis of cell activity during intraoperative surgery[11]. We explored the dynamic behaviour of cancerous cells in mouse intestine. The sample was extracted from the small intestine mucosa and imaged with FFOCT. By assessing the morphology of this wide field image (3800x2560µm$^2$), we were able to locate the border of the cancerous area (Figure 5_B1) and then select 3 fields of 800x800µm$^2$ inside the cancerous zone (Figure 5_B2), in the healthy tissue (Figure 5_A2) and a last one at the border (Figure 4_C1). From these images we can clearly identify a different contrast between normal and pathological tissue. The healthy tissue presents two types of cell. The morphology of these cells and their localization correlate well with histology of similar tissues[20]. At the periphery of the villosities are columnar epithelium cells that are larger and more compact than enterocyte cells which are in a disorganized pattern. In the cancerous zone, the FFOCT image shows highly disorganized fibrous structures, while D-FFOCT produces a complementary image with only cells visible, without the stationary collagen fibres, giving the image a contrast similar to two photon microscopy[21]. We identify a cancer nodule in this zone surrounded by intense signals from cells present all over the cancerous part (Figure 5_D2). These cells are associated with cancerous tissues and were found in every cancerous sample we studied. What we believe to be immune cells are also identified at high density in these tissues. The same cells were recognized in spleen, which is known to have a central function in the immune system[22]. In recent studies we had the opportunity to image human breast cancer tissue where the same cells were present at limits between normal and cancerous tissues.

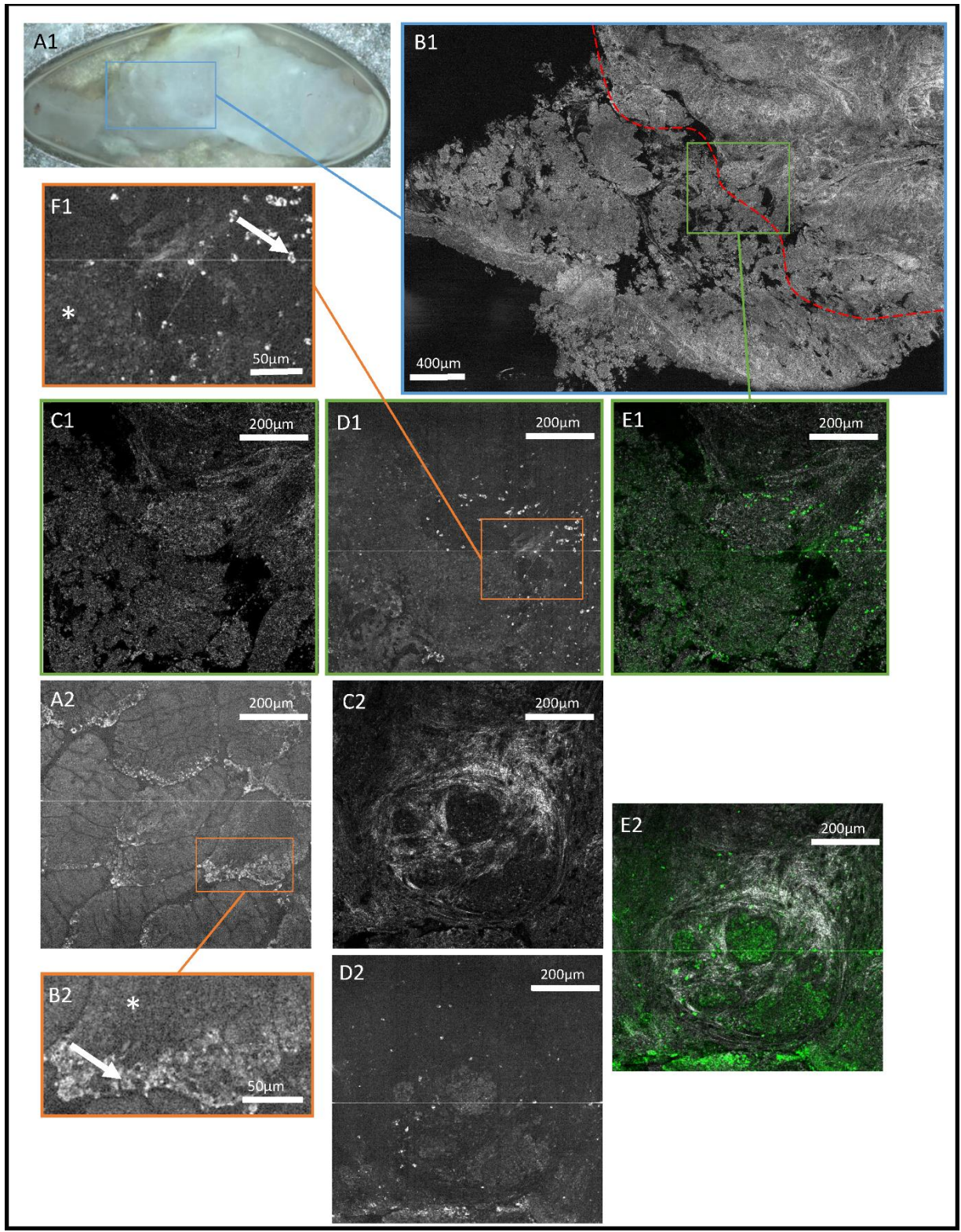

Figure 5: (A1) is the macroscopic photograph of the whole intestine sample. (B1) is a wide field FFOCT image (3800x2560µm$^2$) of the blue zone identified in (A1), the red line is a rough delimitation between the cancerous area and the healthy tissue. (C1) and (D1) are respectively FFOCT and D-FFOCT zooms of the green zone identified in (B1). (E1) is a composition of (C1) in grey and (D1) in green. (F1) shows the details of a border in (D1), the asterisk indicates normal intestine cells while the arrow points towards what we believe to be an immune cell.

(A2) D-FFOCT image of healthy intestinal tissue in mouse with a zoom in (B2) where two types of cells are pointed out: enterocytes (asterisk) and columnar epithelium (arrow). (C2) is a FFOCT image of cancerous intestinal tissue in mouse. (D2) is the corresponding D-FFOCT image of the same field and (E2) the composition of (C2) in grey and (D2) in green.

## 5 Conclusion and perspectives

We have demonstrated that D-FFOCT can provide complementary subcellular contrast to FFOCT images. As FFOCT has already been shown to have good sensitivity and specificity for cancer diagnosis[11,12], we anticipate that revealing more cellular detail including information on cell activity will add to FFOCT's value as an intraoperative tool. The images described here were obtained with a standard FFOCT setup, with simple additional processing to reveal the D-FFOCT signal. The recording takes 1 to 10 seconds and thus could be fully compatible with intraoperative surgery requirements.

Moreover, D-FFOCT has potential for numerous applications in the medical field. Pathologies could be studied and the action of drugs assessed using this method on realistic 3D tissue models or on small animals. In this way, pharmaceutical research could benefit from direct measurement of the effect of drugs on both morphology and metabolism. Immune cell density in cancer following treatment could also be an interesting application. For instance, during surgery to remove a tumour, identification of the tumour margins is crucial and difficult. Margins are currently identified intraoperatively by eye and by frozen section, followed by post-operative histology. The whole histology process gives a diagnosis under one to two weeks. If histology shows presence of remaining cancer, the patient is called back for repeat surgery. The D-FFOCT diagnosis coupled with FFOCT could potentially replace the current intraoperative diagnosis and so reduce the number of patient call-backs after histology. Tissue banking could benefit from the D-FFOCT diagnosis for assessing the quality of the stored samples and select the healthiest ones for transplants or grafts.

The preliminary results presented here suggest a number of improvements to the setup and signal processing. The setup is currently being upgraded with a new CMOS camera (Adimec, Netherlands) with a 10 times larger full well capacity, a 5 times higher frame rate and twice the number of pixels than the camera used for this study. The use of this new camera will increase both the signal to noise ratio and the timescale range that could be explored.

Temperature has a significant effect on cell metabolism; activity is known to drop as the temperature deceases. Building a system capable of maintaining the sample at 37°C should be possible by controlling the temperature gradients. Moreover it could be useful to maintain the tissue being studied in an oxygenated state with a glucose supply.

Among the various possibilities that were explored to process and display the signal, singular value decomposition (SVD) appears promising as it mixes space and time dependant signals[23], and is already used in our laboratory for Doppler ultrasonic imaging. Working with megapixels and time series of 1000 images, we found calculations took too long for intraoperative use. For such applications, one would benefit from the use of GPU to parallelize the processing stage. Cell classification could be attempted by frequency filtering. Useful diagnostic parameters could be extracted by taking advantage of the full autocorrelation function.

## Acknowledgement

We would like to thank the ESPCI neurobiology team of Zsolt Lenkei, especially Diana Zala and Maureen McFadden for animal tissues and advice on biological questions, and Danijela Vignjevic's team at the Institut Curie, in particular Ralitza Staneva for studies on cancer in small animal models. We thank all the LLTech staff for their support, the Curie Hospital and the Institut Gustave Roussy for allowing us to make preliminary tests on human tissues, Amaury Badon for the preliminary tests on SVD and for useful advice for future development, Amy L. Oldenburg for communicating her preprint before publication and finally Kate Grieve for her time and valuable suggestions.